\documentclass[aps,twocolumn,showpacs,floatfix,superscriptaddress,amsmath,amssymb,prl]{revtex4-1}
\usepackage{amsmath}
\usepackage{amsthm}
\usepackage{amssymb}
\usepackage{revsymb}
\usepackage{bbold}
\usepackage{graphicx,color}
\usepackage{babel}
\usepackage{bm}
\usepackage{hyperref}



\newcommand{\ad}{\mathop{\text{ad}}\nolimits}

\definecolor{dgreen}{rgb}{0,0.5,0}

\definecolor{delete}{cmyk}{0.5,0,0,0}

\begin{document}
\title{Quantum Distance to Uncontrollability and Quantum Speed Limits}
\author{Daniel Burgarth}
\address{Center for Engineered Quantum Systems, Dept. of Physics \& Astronomy, Macquarie University, 2109 NSW, Australia}
\author{Jeff Borggaard}
\address{Interdisciplinary Center for Applied Mathematics and the Department of Mathematics at Virginia Tech, Blacksburg, VA 24061, USA.}
\author{Zolt\'an Zimbor\'as}
\address{Wigner Research Centre for Physics, H-1121, Budapest, Hungary}
\address{MTA-BME Lend\"ulet Quantum Information Theory Research Group Budapest, Hungary} \address{Mathematical Institute, Budapest University of Technology and Economics,\\ H-1111, Budapest, Hungary}
\begin{abstract}
Distance to Uncontrollability is a crucial concept in classical control theory. Here, we introduce Quantum Distance to Uncontrollability as a measure how close a universal quantum system is to a non-universal one. This allows us to provide a quantitative version of the Quantum Speed Limit, decomposing the bound into a geometric and dynamical component. We consider several physical examples including globally controlled solid state qubits and a cross-Kerr system, showing that the Quantum Distance to Uncontrollability provides a precise meaning to spectral crowding, weak interactions and other bottlenecks to universality. We suggest that this measure should be taken into consideration in the design of quantum technology.
\end{abstract}
\maketitle
\emph{Introduction}-- Just as it is in our day-to-day computers, \textit{universality} - the ability to run any algorithm in principle- is the central concept in quantum computing. In the current race to prove the first traces of it, and with the first success to report it in larger systems \cite{Arute2019}, this is more true than ever. It is often argued \cite{Lloyd1995} that universality itself is \textit{universal}, e.g. that almost all systems are universal, and if not, a slight change of parameters would render them so. This is even true in noisy systems, where universality needs to be combined with error correction.

However, we argue, that there is a flip-side to this: if any non-universal system is close to a universal one, then also many universal ones are dangerously close to non-universal ones. Universality might be unstable or inefficient then. Indeed, it seems that nature is hesitant to explore high-dimensional dynamics \cite{Poulin2011} and simple non-universal systems are often good approximations. Experimentalists working hard to engineer weak non-linearities in quantum optics, weak anharmonicities in superconducting systems, or to avoid spectral crowding in solid state systems are well aware of such limitations. Here, we put  this intuition in a precise framework we call \textit{quantum distance to controllability}, and we show how it relates to a notoriously difficult to compute yet independently interesting quantity: the \textit{quantum speed limit} \cite{Deffner2017, Caneva2009,Lee2018}. Intuitively speaking, the distance to uncontrollability identifies and quantifies the \emph{control bottleneck} of a quantum system.

But first, let us put our result into context. Universality is also universal in (Kalman) linear control, a key subject in engineering with a wide range of applications ranging from mobile communications to space travel. The concept of distance to controllability was introduced in the linear setting \cite{paige1981properties,eising1984between} to quantify the smallest perturbation that would lead to an uncontrollable system. There is no no speed limit in linear control, and distance to controllability is used as a test of the numerical robustness of controllability.

In quantum mechanics, our results rely on the simple observation that if a controllable quantum system is close to an uncontrollable one, nature needs time to distinguish them. This should provide us with a bound for the quantum speed limit. In order to make this precise we have to find a path-independent controllability criterion, and solve a group theoretic worst-case scenario. We solve both using very recently developed techniques and insights on controllability \cite{Zeier2011,Zeier2015}.

The latter is worthwhile motivating independently. Let us imagine Bob wants to buy a quantum computer, but unfortunately the shop is out of stock of universal machines. However, he can choose amongst \textit{any non-universal device}. Bob knows that his friend Alice will try to run the \textit{hardest possible} algorithm on whichever machine he chooses. Which is the best machine to buy, and how much is his machine going to fail at running Alice’s task? We give a dimension-independent bound to this question.

\emph{Setup}-- We consider a control system given by a time-dependent Schr\"odinger Equation
\begin{equation}
	\frac{d}{dt}U(t)=i H(t) U(t),
\end{equation} 
where $U$ and $H$ are complex valued square matrices of size $d<\infty$ and $U(0)=\mathbb{1}$. First, for simplicity, we assume that $H(t)$ decomposes into a drift and a single control as
\begin{equation}
	H(t)=H_d + f(t) H_c,
\end{equation}
with $H_d, H_c$ self-adjoint and traceless and $f(t)$ arbitrary piece-wise continuous. The extension to multiple controls and other classes of control pulses will be considered later. It is well known that the pair $(H_d,H_c)$ is controllable if and only if the smallest real Lie algebra $\mathfrak{g}$ that contains $iH_d$ and $iH_c$ and has maximal dimension,
\begin{equation}
\dim \mathfrak{g} (H_d,H_c) =d^2-1.
\end{equation}
From now on we assume that $(H_d,H_c)$ is such a controllable pair. In this case there is a minimal time $T_*>0$ such that \emph{all} unitaries $U$ can be reached \emph{exactly} at that time \cite{Jurdjevic1972}. We refer to this as the \emph{control time} and note that not much is known about the dependence of the control time on the drift and controls. Interestingly in linear control theory controllable systems can, in principle, be controlled arbitrarily fast. In our case, however, it is clear that even for unbounded controls $f(t)$ the evolution of the drift sets a `quantum speed limit' (we first consider pulses as `free resources' and later look at more general scenarios with additional constraints). The goal of the present study is to understand this limit better by finding the distance to uncontrollability. 

More specifically, consider the smallest self-adjoint modification to the drift which renders the system not fully controllable (here referred to as `uncontrollable'):
\begin{equation}
\label{epsilon} \varepsilon_*  = \inf \{ \| \Delta H \|:\dim \mathfrak{g}(H_d{+}\Delta H,H_c)<d^2{-}1 \}
\end{equation}
where $\Delta H=\Delta H^\dagger$ and $\|A\|\equiv \|A\|_\infty$ the operator norm . Notice that $(0,H_c)$ is uncontrollable ($\dim \mathfrak{g}(0,H_c)=1$) and therefore $\varepsilon_*\leq \| H_d\|$.
If a system is uncontrollable, intuitively there is a least one `direction' in the dynamics missing. This will be made more precise later. Let us denote with $\mathcal{R}$ the set of unitaries that \emph{can} be reached in the not fully controllable system  $(H_d+\Delta H,H_c)$. We are then interested in
\begin{equation}
	\delta(\mathcal{R})=\sup_{U\in SU(d)}\inf_{V\in\mathcal{R}}\|U-V\|, 
\end{equation}
which describes the unitary that has the worst approximation within $\mathcal{R}$.
For any uncontrollable system $(H_d+\Delta H,H_c)$ there is a $U$ such that any evolution is at least at distance $\delta(\mathcal{R})$ to $U$:
\begin{equation}
	\|\mathcal{T}e^{i\int_0^{T}(H_d+\Delta H+f(t) H_c)dt}-U\|\ge \delta(\mathcal{R})
\end{equation} for \emph{any} time $T$ and \emph{any} control pulse $f(t)$ on $[0,T].$
 On the other hand, by the definition of $T_*$ there is a control pulse $f$ on the interval $[0,T_*]$ such that
\begin{equation}
	\mathcal{T}e^{i\int_0^{T_*}(H_d+f(t) H_c)dt}=U.
\end{equation}
A simple calculation \cite{Nielsen2006a,Arenz2017d} using the unitary invariance of the operator norm states that for this $f(t)$ we have
\begin{equation}
	\|\mathcal{T}e^{i\int_0^{T_*}(H_d+\Delta H+f(t) H_c)dt}-U\|\le T_* \|\Delta H\|.
\end{equation}
Since this holds for any $\Delta H$ rendering the system uncontrollable we get
\begin{equation}
	\left. T_*\ge \delta(\mathcal{R}) \middle/ \varepsilon_*\right..
\end{equation}
This bound characterises the speed limit in terms of a geometric term $\delta(\mathcal{R})$ and a dynamical one $\epsilon_*$. 

Note that  $\delta(\mathcal{R})$ still depends on the geometry of specific control system, and computing $\mathcal{R}$ typically involves a Lie closure. For applications in quantum information theory $\mathcal{R}$ can be exponentially large, and therefore $\delta(\mathcal{R})$ is hard to obtain. 
Thus, it is interesting to introduce a universal bound \begin{equation}
		\delta_*=\inf_{\mathcal{R}\neq SU(d)}\delta(\mathcal{R}),
	\end{equation}
	which is independent of the details of the model.
	This allows us to obtain a non-trivial bound on the control time,\begin{equation}
	\left.	T_*\ge \delta_* \middle/ \varepsilon_*\right..
	\end{equation}
 The next part of this manuscript is dedicated to finding explicit expressions on the geometric $\delta_*$ (which is only a function of the dimension $d$ of Hilbert space -- we however give a dimension independent lower bound) and on the dynamical quantity $\varepsilon_*=\varepsilon_*(H_d,H_c)$. We will then look at examples.

\emph{Understanding the geometric component $\delta_*$}--
The simplest case is when $\mathcal{R}$ has a symmetry, that is, there is a non-trivial projection $P$ such that $[\mathcal{R},P]=0$. In such a case, we may choose a target unitary $U$ which moves a particular state $|\psi\rangle$ in the range of $P$ to an orthogonal state in its kernel. Then, for all $V\in\mathcal{R}$, $\|V-U\|\ge \|(V-U)|\psi\rangle\|=\sqrt{2}$ (the distance between two orthogonal states). As an example this provides the loose bound
\begin{equation}
   \left. T_*\ge \sqrt{2} \middle/ \|H_d\| \right.,
\end{equation} which is an instance of \cite{Arenz2017d}. In general, this approach does not work, since there are uncontrollable systems without symmetry (see the first physical example below). 

In the supplementary material, we use recent results \cite{Zeier2011,Zimboras2015} that show uncontrollable systems always have non-trivial symmetries on a doubled system. This allows us to make an argument similar to the simple case above. However, in general even fully controllable systems in the doubled space cannot move states in the range of such symmetries to their kernel, and we need to use a group averaging argument to see how far we can move them at least. Both steps (doubling and averaging) cost us some constants in the bound, but we show that $\delta_*\ge 1/4$. This is remarkable, because it is independent of the dimension, and only a factor of $8$ from the trivial bound $\delta_*\le 2$. In terms of the quantum speed limit, it means we can eliminate the geometric component at little cost and obtain the bound
\begin{equation}
	\left.	T_*\ge 1 \middle/ 4 \varepsilon_*\right.,
	\end{equation} which should be considered as a milestone result of this manuscript.
	
 \emph{Understanding the dynamical component $\varepsilon_*$}-- At first glance $\varepsilon_*$ is daunting. One reason is the Lie criterion. As opposed to linear systems, where controllability can be checked by computing a the rank of the controllability matrix with \emph{fixed dependence} on the perturbation, here we compute the rank of certain Lie polynomials which depend on the generators. While perturbing the system with $\Delta H$ can change this rank, this does not imply uncontrollability: it could just mean that one has to construct another set of polynomials. Fortunately there is a recent, powerful alternative characterisation of controllability which circumvents this problem \cite[Theorem 21]{Zeier2011}:
	\begin{align}
	&	\dim \mathfrak{g} (H_d+\Delta H, H_c) <d^2-1 \Leftrightarrow \nonumber \\
	& \dim \{(H_d+\Delta H)^{(2)}, H_c^{(2)}\}'>2,
	\end{align}
	where $A^{(2)}\equiv A\otimes \mathbb{1}+\mathbb{1}\otimes A$ denotes the tensor symbolisation of a matrix $A$ on the doubled space, and $\{S\}'$ is the commutant of the set S (the vector space of matrices commuting with all elements of $S$). We can boil this down to something more standard by noting that $X$ commutes with a matrix $B$ if and only if $B^{(\ad)}\operatorname{vec}(X)\equiv (B\otimes \mathbb{1} - \mathbb{1}\otimes B^T)\operatorname{vec}(X)=0$, where we used row vectorisation. The dimension of the commutant becomes equivalent to the nullity of the $2d^4\times d^4$ matrix
\begin{equation}
R(H_d+\Delta H,H_c) \equiv 	\begin{pmatrix}
		\left(i (H_d+\Delta H)^{(2)}\right)^{(\ad)}\\
		\left( i H_c^{(2)}\right)^{(\ad)}
	\end{pmatrix}.
\end{equation}
Finally controllability is equivalent to this matrix having a rank of $d^4-2$ and $\varepsilon_*$ can be defined by reducing this rank with a minimal Hermitian choice of $\Delta H$:
\begin{equation}\label{structured}
\varepsilon_* = \inf \{ \| \Delta H \|:\operatorname{rank}R(H_d{+}\Delta H,H_c)<d^4{-}2  \}.
\end{equation}
Having mapped Eq. (\ref{epsilon}) to a rank problem, we may use the vast literature on algorithms designed to approximate $\varepsilon*$ in the classical case. For recent computable bounds see \cite{ebihara2006estimates} and in particular for structured (including Hermitian) perturbations see \cite{Karow2009}.   Because the rank criterion (\ref{structured}) involves a structured optimisation not considered in the previous literature, we were only able to find \emph{lower} bounds on $\epsilon_*$ with the classical algorithms. We therefore discuss two alternative approaches for the quantum case: one based on energy gaps, and one on graph theory.
	
	\emph{Relating $\epsilon_*$ to the energy gap}-- There is an intricate relationship between symmetry and controllability \cite{Zeier2011,Wang2010,Zimboras2015}. Altafini \cite{Altafini2002} and Turinici \cite{Turinici2003} considered sufficient criteria based on the absence of certain degeneracies. Here we make the following simple observation. Assume the drift $H_d$ has an n-fold degeneracy, e.g. there are orthogonal eigenstate $\{H_d|e_k\rangle=e|e_k\rangle,k=1,\ldots,n\}$. Assume further that the control acts only non-trivially on a subspace $\mathcal{H}_c$, that is, $\mathcal{H}=\mathcal{H}_c\oplus\mathcal{H}_c^\perp$, and that $\mathcal{H}_c$ is left invariant by the control. Then if $\dim \mathcal{H}_c<n$, there is a symmetry in the control system, $\exists M=M^\dagger: [H_c,M]=[H_d,M]=0$, and $\delta_*\ge \sqrt{2}$. To see this, simply build linear combinations of degenerate eigenstates (of which there are enough) to get an eigenstate of $H_d$ that has no support on $\mathcal{H}_c$. The corresponding projector will be such a symmetry, and by linearity and the Jacobi identity, it will commute with $\mathcal{R}$ and thereby render the system not fully controllable. To summarise: by looking at the block structure of the control(s), and the spectrum of the Hamiltonian, we can see how to make the system uncontrollable. In particular, when $H_c$ is rank $1$, we get $\epsilon_*<\min \Delta e_k,$ the minimum gap of the energy eigenstates of $H_d$.	

\emph{Relating $\epsilon_*$ to graph properties and min-cut}-- We bring $H_c=\sum_k |e_k\rangle \langle e_k|$ into a diagonal representation. If $H_0+\Delta H$ is not a connected graph in this basis, the system is not controllable \cite{Altafini2002}: the system splits into a block structure with a corresponding symmetry, and $\delta_*\ge \sqrt{2}$. To find such a perturbation efficiently, we can consider a graph with weights $|\langle e_k |H_0|e_j\rangle|$ and run the polynomial sized  Stoer-Wagner algorithm \cite{Stoer1994a}. It provides the minimal $\Delta H$ in the $L_{1,1}$ norm, and thereby an interesting upper bound on $\epsilon_*$.
Before we consider examples, let us also generalise to multiple control functions that are possibly bounded.

\emph{Bounded \& multiple controls}-- 
Often, control amplitudes are bounded, and in such case we can compute a distance to uncontrollability also by perturbing the controls. We may consider a time-dependent Hamiltonian $H(t)=\sum_{j=1}^M g_j(t)\tilde{H}_j +\sum_{k=1}^L f_k(t)H_k$  with the first set of controls bounded as $|g_j(t)|<c$ and the remaining controls not bounded. This notation can also include a drift term with the bound $g(t)=1$. If now the generators $(\tilde{H}_1,\cdots ,\tilde{H}_M,H_1,\cdots,H_L)$ are controllable, and the modified set $(\tilde{H}_1+\Delta_1,\cdots ,\tilde{H}_M +\Delta_M, H_1,\ldots, H_L)$, $\|\Delta_j\|\leq \epsilon_*/M$ is uncontrollable, we can derive a similar bound as above. For $U_1$ generated by $\sum_{j=1}^M g_j(t)\tilde{H}_j+\sum_{k=1}^L f_k(t)H_k$ and $U_2$ generated by $\sum_{j=1}^M g_j(t)(\tilde{H}_j+\Delta_j) +\sum_{k=1}^L f_k(t)H_k$ on $[0,T_*]$ we have $\|U_1-U_2\|\leq \int_0^{T_*} \|g(t)\Delta \|\leq c \epsilon$
and therefore $T_*\geq \delta_*/(c\epsilon_*)$. Having generalised to this more realistic scenario, we now look at four paradigmatic examples.

\emph{Physical Examples}-- 
As a first and simple example consider two qubits with full local control and the drift $\delta Z\otimes Z$, where $Z$ denotes a Pauli matrix. Even when uncontrollable, this system has no symmetry, and we use $\delta_*\ge 1/4$.  Our bound provides $T_* \ge 1/(4\delta)$ while the exact limit was shown to be \cite{Khaneja2001} $T_*=\pi/(2 \delta)$, which only differs from our bound by a constant factor.

\emph{Global controls}--  This is a commonly encountered situation in solid state quantum technology. The system with drift $H_d=\sum Z_i\otimes Z_j$ and global controls $g_1(t)\sum \gamma_i X_i$ and $g_2(t)\sum \gamma_i Y_i$ is fully controllable as long as $|\gamma_i|\neq |\gamma_j|$ \cite{Albertini2002}. Therefore, $\epsilon_*\le \Delta \gamma \equiv  \min_{ij}(|\gamma_i|-|\gamma_j|)$. The corresponding quantum speed limit with bounded controls $g_i(t)\le c$
\begin{equation}
   \left. T_*\ge \sqrt{2}\middle/ c \Delta \gamma\right.
\end{equation} makes spectral crowding \cite{Schutjens2013a} rigorous.

\emph{Local controls}--	As our third example, we consider the nearest-neighbour control system given by $ H_d=\sum_{n=1}^{d-1} |n \rangle \langle n+1| +|n+1 \rangle \langle n|$ and 
	$H_c=|1\rangle\langle 1|$ on a $d$-dimensional space with orthonormal basis $| n\rangle$.  This is one of the gold standards \cite{Lee2018,Banchi2017} in Quantum Speed Limits. The control is rank $1$ so we can use the energy gap to bound the speed limit. The spectrum of the Hamiltonian is simply given by $e_k=2\cos\left(\frac{k\pi}{d+1}\right),\ k=1,\dots,d$. The minimum gap can be upper bounded by $3 \pi^2/d^2$ and this bound becomes tight as $d\rightarrow \infty$. Therefore, we can find a lower bound on the control time as 
	\begin{equation}
	\left.	T_*\ge\sqrt{2} d^2\middle/ 3\pi^2 \right..
	\end{equation}
To our knowledge this is the first analytical proof that control times scale at least quadratically with the dimension in this system. 

\emph{Cross-Kerr interaction with fixed particle number}-- Linear optics, both passive and active, is a premier platform of quantum information processing \cite{Weedbrook2011}. Thus, the problem of extension to universality of passive linear optics has gained a lot of attention in recently. 
In particular, it has been shown
that adding a cross-Kerr interaction leads to universality \cite{Oszmaniec2017, Lloyd1999}. This is of relevance in quantum metrology with random bosonic states \cite{Oszmaniec2016}.

Here we will perform our analysis of distance to controllability in such a platform, considering $d$ qumodes with $N$ photons.
The Hamiltonians that we can implement are any linear optical gates with essentially arbitrary strength, and cross-Kerr gates between nearest-neighbor modes with a strength that is typically strongly bounded, i.e., we have a Hamiltonian of the form:
\begin{equation}
    H= \sum_{j=1}^{d-1} g_j(t) n_jn_{j+1} + \sum_{k \le l}^{d} f_{k,l}(t) a^\dagger_k a^{\phantom{\dagger}}_l
    {+} h.c.,
\end{equation}
where $a^\dagger_j$ and $a_j$ are the creation and annihilation operators and $n_j= a^\dagger_ja_j$ is the particle number operator corresponding to mode $j$. The control functions $f_{k,l}$ can be considered unbounded, while the control functions for the cross-Kerr interactions are bounded $g_j(t) \le c$.  We can now upper-bound the $\epsilon_*$  by choosing $\Delta  =  \sum_j n_j n_{j+1}$. One can easily show that $\| \Delta \| =  \frac{N^2}{4}$. Thus we arrive at
\begin{equation}
  \left.  T_* \ge 1\middle/ c N^2\right.,
\end{equation}
which shows us how gate times scale with the number of photons and the strength of the cross-Kerr interaction.

\emph{Conclusions}-- We have introduced the quantum distance to uncontrollabilty as a quantitative measure how good a control systems is, and derived bounds on the quantum speed limit. These are provided in terms of a dynamical and geometric component. We obtained a dimension-independent bound for the geometric component, and linked the computation of the dynamical part to energy gaps and to graph theory, which provides an easy route to computing speed limits. We gave several physical examples from quantum technology. Maximizing the distance to the nearest uncontrollable system has also been used in the design of linear controlled systems~\cite{borggaard2011using}, and it would be interesting to consider it in the design of quantum computers.

\emph{Acknowledgements}-- DB acknowledges discussions with Mattias Johnsson about quantum speed limits. This research was funded in part by the Australian Research Council (project number FT190100106) and the National Science Foundation (grant number DMS-1819110). ZZ would like to acknowledge support from the NKFIH Grants No. K124152, K124176, KH129601, K120569 and from the Hungarian Quantum Technology National Excellence Program Project No. 2017-1.2.1-NKP-2017-00001.
\section{Supplementary Material}
Here we prove that $\delta_*\ge 1/4$. First, let us derive some simple norm bounds. Note that we can bound norm differences on a doubled system as
\begin{align}
& \| U_1\otimes U_1 - U_2 \otimes U_2 \| = \nonumber \\
& \|(U_1\otimes U_1 - U_1 \otimes U_2) + (U_1\otimes U_2 -U_2 \otimes U_2)\| \nonumber\\
& \le \|U_1 - U_2\| \|U_1\| +   \|U_1 - U_2\| \|U_2\|\nonumber \\
& \le  2\|U_1 -U_2 \|.
\end{align}
Next, we can relate norm differences of unitaries to their action on states as follows. \begin{align}& \|V \rho V^\dagger - W \rho W^\dagger  \|_1\nonumber\\ & = \| \rho - V^\dagger W \rho (V^\dagger W )^\dagger  \|_1 \nonumber\\ & = 
		 \| \rho - V^\dagger W \rho + V^\dagger W \rho - V^\dagger W \rho (V^\dagger W )^\dagger  \|_1\nonumber \\ &  \le \| \rho - V^\dagger W \rho  \|_1 +  \|V^\dagger W \rho - V^\dagger W \rho (V^\dagger W )^\dagger  \|_1 \nonumber\\ & = 2 \| (1 - V^\dagger W ) \rho\|_1.\end{align}  Using H\"older's inequality, we get:
		\begin{align}&\|V \rho V^\dagger - W \rho W^\dagger  \|_1 \nonumber \\ &\le 2 \| \rho \|_1  \|(1 - V^\dagger W ) \|_{\infty} =  2 \|V-W \|,\end{align} where we dropped the $\infty$ subscript from the operator norm.

	Now suppose $\mathcal{R}\neq SU(d).$ Then by compactness of $SU(d)$, the closure $\overline{\mathcal{R}}\equiv G$ is a connected and \emph{strict} subgroup of $SU(d)$. By the above, 
		\begin{align} & \delta(\mathcal{R})=\sup_{U\in SU(d)}\ \inf_{V\in \mathcal{R}} \| V - U \| \nonumber \\ & \ge \sup_{U\in SU(d)}\ \inf_{V\in G} \| V - U \| \nonumber \\ & \ge   \frac{1}{2} \sup_{U\in SU(d)} \inf_{V\in G}\| V \otimes V- U \otimes U\|\nonumber \\ 
		& \label{mega} \ge  \frac{1}{4}\sup_{U\in SU(d)} \inf_{V\in G}  \| \textrm{Ad}_{V \otimes V} (\rho)  {-}  \textrm{Ad}_{U \otimes U} (\rho)   \|_1 ,
	    \end{align} where $\rho$ is an arbitrary state on the doubled system and $\textrm{Ad}_{W} (\rho) = W \rho W^\dagger$.
		
		We will now use a representation theory argument laid out in \cite{Zeier2011,Zeier2015,Zimboras2015} to construct a state $\rho$ that is left invariant by $V\otimes V$. Consider $G^{\otimes 2}=\{g\otimes g |g\in G\}$.	It is known that this representation of $G$  has at least two invariant subspaces, the symmetric and the antisymmetric subspace. It also follows from  \cite[Theorem 6]{Zeier2015} that because $G \ne SU(d)$, then either the symmetric or the antisymmetric subspaces of $G^{\otimes 2}$ (or both) break into more than one irreps of $G$. Let us assume it happens in the symmetric space, with the other case treated analogously. Thus there is a projection $P$,  such that for any $V \in G$, we have $V\otimes V P (V\otimes V)^\dagger= P$; and $P P_+ = P$, where $P_+$ is the projection to the symmetric space, and  $P\neq P_+$.
	    We can assume that the dimension (rank) of $P$ is less than or equal to half of that of $P_+$ (otherwise consider its compliment in $P_+$). Since projections are positive, we can normalise them into a state $\rho=P/\dim P$ that will be a good choice for Eq.~(\ref{mega}).
	    
	    Due to its invariance, the infinum in  Eq.~(\ref{mega}) disappears and we obtain
	    \begin{equation}  \delta(\mathcal{R}) \ge  \frac{1}{4}\sup_{U\in SU(d)}   \|  \rho    -   U \otimes U \rho  (U \otimes U)^\dagger \|_1 .\end{equation} We now need to find a $U\in SU(d)$ which moves $\rho$ far away. This would be a hard problem in general, but we can use Jensen's inequality and move to the average case, which is good enough:

	    \begin{align}  \delta(\mathcal{R}) &\ge  \frac{1}{4}  \|  \rho    - \int_{SU(d)}  (U \otimes U) \rho  (U \otimes U)^\dagger \|_1\nonumber\\
	    & \ge \frac{1}{4}\| P/\dim P   -  P_+/\dim{P_+} \|_1 \ge 1/4.\end{align}
	    In the second step, we used that $SU(d)$ acts irreducibly in $P_+$. In the last step, we went to a joint diagonal representation of $P$ and $P_+$ and used the fact that $\dim P \le \dim P_+/2$ to explicitly bound the trace norm.

\bibliographystyle{apsrev4-1}
\bibliography{referencesJeff,danielreferences}
\end{document}